\documentclass[prl,twocolumn,superscriptaddress,preprintnumbers,nofootinbib]{revtex4-2}

\usepackage{graphicx}
\usepackage[mathlines]{lineno}

\usepackage{times}
\usepackage{dcolumn}
\usepackage{bm}
\usepackage{hyperref}
\usepackage{subfigure}
\usepackage{float}
\usepackage{wrapfig}
\usepackage{amssymb,latexsym, amsmath}
\usepackage{graphics}
\usepackage{float}
\usepackage{appendix}
\usepackage{amssymb}
\usepackage{latexsym,amsthm,makeidx,enumerate}
\usepackage{mathrsfs}
\usepackage{siunitx}
\usepackage{comment,hhline,multirow,setspace}

\usepackage{dcolumn}
\usepackage{bm}
\usepackage[T1]{fontenc}
\usepackage{amsmath}
\usepackage[compact]{titlesec}
\usepackage{graphicx}
\usepackage{tabularx}
\usepackage{mathtools}
\usepackage{xfrac}
\usepackage{orcidlink}
\usepackage{natbib}
\usepackage{hyperref}\hypersetup{colorlinks=true, linkcolor=blue, citecolor=blue, urlcolor=blue, pdftitle={}, pdfauthor={}}
\usepackage{lineno}

\newcommand{\WSe}{WSe$_{2}$ }
\newcommand{\WTe}{WTe$_{2}$ }

\newcommand{\WSeTe}{W[Te$_x$Se$_{1-x}$]$_{2}$ }
\newcommand{\WSeTed}{W[Te$_x$Se$_{1-x}$]$_{2(1-\delta)}$ }
\graphicspath{{graphics/}}

\usepackage{setspace}
\begin{document}

\makeatletter
\let\NAT@bare@aux\NAT@bare
\def\NAT@bare#1(#2){%
	\begingroup\edef\x{\endgroup
		\unexpanded{\NAT@bare@aux#1}(\@firstofone#2)}\x}
\makeatother

\makeatletter
\newcommand{\printfnsymbol}[1]{%
  \textsuperscript{\@fnsymbol{#1}}%
}
\DeclareSIUnit{\angstrom}{\textup{\AA}}

\title{\Large 
{\color{red}}Defect-induced multiferroicity in bulk solid solutions of WSe$_2$ and WTe$_2$}
\author{H. Rojas-Páez}
\affiliation{Department of Physics, Universidad de Los Andes, Bogot\'a 111711, Colombia}
\author{G. Villabón-Linares}
\affiliation{Department of Physics, Universidad de Los Andes, Bogot\'a 111711, Colombia}
\author{J. Pazos}
\affiliation{Faculty of Engineering and Basic Sciences, Universidad Central, Bogot\'a, Colombia}
\author{E. Ramos}
\affiliation{Department of Physics, Universidad de Los Andes, Bogot\'a 111711, Colombia}
\author{R. Moreno}
\affiliation{Department of Physics, Universidad de Los Andes, Bogot\'a 111711, Colombia} 
\author{O. Herrera-Sandoval}
\affiliation{Faculty of Engineering and Basic Sciences, Universidad Central, Bogot\'a, Colombia}
\author{J. A. Galvis\orcidlink{0000-0003-2440-8273}}
\affiliation{School of Engineering, Science and Tecnology, Universidad del Rosario, Bogot\'a, Colombia}
\author{P. Giraldo-Gallo$^{\dagger}$\orcidlink{0000-0002-2482-7112}}
\affiliation{Department of Physics, Universidad de Los Andes, Bogot\'a 111711, Colombia}

\date{\today}
\begin{abstract}
Transition-metal dichalcogenides provide a versatile platform for tunable ferroic phenomena at the atomic scale owing to their reduced dimensionality. Here, we investigate the structural, magnetic, and ferroelectric properties of bulk solid-solution $\mathrm{W(Se_{1-x}Te_x)_{2(1-\delta)}}$ single crystals synthesized by chemical vapor transport. The room-temperature behavior is analyzed as a function of tellurium concentration ($x$) and chalcogen defect fraction ($\delta$). X-ray diffraction and Raman spectroscopy reveal lattice expansion and symmetry reduction with increasing $x$, consistent with a $2H \rightarrow 1T_d$ structural transition above a critical composition $x_c \approx 18\%$. Piezoresponse force microscopy identifies piezoelectricity near stoichiometric compositions ($|\delta|<5\%$) and switchable ferroelectricity in the chalcogen-deficient regime ($\delta>20\%$). Magnetometry measurements show a corresponding evolution from paramagnetic to ferromagnetic behavior with increasing $\delta$. Near-stoichiometric, Te-poor samples exhibit piezoelectric and paramagnetic responses, whereas multiferroic states—characterized by the coexistence of ferroelectric and ferromagnetic responses—emerge at high vacancy concentrations. The performed characterizations indicate that $x$ primarily governs structural symmetry, while $\delta$ controls the emergence of both ferromagnetic and ferroelectric responses. These trends are summarized in a configurational phase diagram highlighting the cooperative influence of dopants and defects on ferroic behavior. Overall, controlled stoichiometry and vacancy engineering offer an effective strategy to tailor ferroic responses in transition-metal dichalcogenides.
\end{abstract}

\maketitle
\def\thefootnote{$\dagger$}\footnotetext{Corresponding author: pl.giraldo@uniandes.edu.co}\def\thefootnote{\arabic{footnote}}

\section{Introduction}\label{introduction}
Multiferroic materials exhibit two or more coexisting ferroic orders—most commonly ferromagnetism, ferroelectricity, and ferroelasticity—within a single phase \cite{spaldin2010multiferroics}. Magnetoelectric multiferroics, which combine ferromagnetic and ferroelectric behavior, are particularly attractive for next-generation data processing and storage, as coupled ferroic orders could enable low-power, non-volatile control of magnetization by electric fields \cite{Fiebig2016evolution}. However, intrinsic multiferroics remain scarce because ferroelectric and magnetic orders usually arise from mutually exclusive electronic configurations: magnetism originates from unpaired $d$ or $f$ electrons, whereas conventional ferroelectricity requires empty $d$ shells to stabilize polar displacements \cite{hill2000there,spaldin2004fundamental}. This fundamental incompatibility between spin and charge asymmetries explains the rarity of intrinsic magnetoelectric coupling in crystalline solids. Alternative mechanisms have therefore been sought to overcome this constraint.

Alternative, non-displacive mechanisms of ferroelectricity have broadened the range of possible multiferroic systems \cite{Fiebig2005revival,barone2015mechanisms}. Type-I multiferroics, in which ferroelectricity and magnetism arise from distinct microscopic origins, include lone-pair–driven (BiFeO$_3$), geometric (YMnO$_3$), and charge-ordering (LuFe$_2$O$_4$) systems \cite{wang2003epitaxial,van2004origin,ikeda2005ferroelectricity,van2008multiferroicity}. In contrast, type-II multiferroics exhibit magnetically induced ferroelectricity, where non-collinear spin textures or exchange striction break inversion symmetry, as in TbMnO$_3$ and Ca$_3$Co$_{2-x}$Mn$_x$O$_6$ \cite{kimura2003magnetic,choi2008ferroelectricity}. Yet most known multiferroics suffer from low ordering temperatures or weak coupling coefficients, motivating the search for alternative routes to robust magnetoelectricity.

Recent evidence shows that crystallographic point defects can stabilize ferroic orders. Defects such as vacancies, antisites, or nonmagnetic adatoms can locally break inversion symmetry, generate magnetic moments, and couple to lattice distortions, enabling ferromagnetism and ferroelectricity in otherwise non-ferroic materials \cite{esquinazi2013defect,esquinazi2020defect}. Theoretical and experimental studies demonstrate that point defects introduce spin-polarized states near the Fermi level, creating localized magnetic moments in systems such as graphite, SrTiO$_3$, and ZnFe$_2$O$_4$ \cite{rata2022defect,zviagin2020control}. In oxides, 2$p$ dopants at oxygen sites can induce spin polarization \cite{bannikov2008magnetism}, while Sr vacancies in SrTiO$_3$ drive tetragonal distortions and ferroelectricity \cite{kang2020ferroelectricity}. Defect dipoles from CuSb$_2$O$_6$ in K$_{0.5}$Na$_{0.5}$NbO$_3$ also tune polarization loops and coercivity \cite{wang2017defect}.

Certain defects further mediate direct coupling between magnetic and electric orders, producing emergent multiferroicity. In YFeO$_3$, antisite defects between Y and Fe atoms break inversion symmetry and induce ferroelectricity within a weakly canted antiferromagnetic structure \cite{ning2021antisite}. Oxygen vacancies likewise play a central role in tuning coupled ferroic behavior in complex oxides \cite{kalinin2013functional,gunkel2020oxygen}. Ordered oxygen-vacancy planes in (LuFeO$_3$)$_9$/(LuFe$_2$O$_4$)$_1$ superlattices stabilize ferrimagnetic and ferroelectric domains simultaneously \cite{hunnestad20243d}, while in nanocrystalline KTaO$_3$, extremely dilute oxygen vacancies and residual Fe impurities (0.04–0.1 mol \%) induce both magnetism and ferroelectricity, which vanish in stoichiometric single crystals \cite{golovina2012defect}. These examples highlight defect engineering as a powerful route to multiferroicity.

In conventional three-dimensional ferroics, the emergence of order parameters is fundamentally constrained by symmetry: noncentrosymmetric crystal structures permit spontaneous polarization, whereas long-range magnetic order is associated with the spontaneous breaking of time-reversal symmetry \cite{aizu1966possible,aizu1970possible}. However, as materials are reduced to the nanoscale, incomplete screening of bound polarization charges gives rise to depolarization fields that destabilize long-range dipolar order, resulting in a critical thickness below which ferroelectricity is suppressed \cite{junquera2003critical,spaldin2004fundamental,nishino2020evolution}. This limitation has stimulated intense interest in two-dimensional (2D) materials, where nonlocal dielectric screening and large surface-to-volume ratios modify Coulomb interactions and enhance sensitivity to local symmetry breaking and defect-induced phenomena \cite{cudazzo2011dielectric,fei2018ferroelectric,hong2015exploring,de2021direct}.

Within this context, layered two-dimensional systems provide a fertile platform for unconventional ferroic and multiferroic phenomena. Intrinsic multiferroicity at reduced dimensionality remains rare, having been reported in only a limited number of systems, including Fe-doped In$_2$Se$_3$ \cite{yang2020iron}, NiI$_2$ \cite{song2022evidence}, CuCrP$_2$S$_6$ \cite{lai2019two}, p-doped SnSe \cite{du2022two}, Ti$_3$C$2$ \cite{tahir2022first}, CuCrSe$_2$ \cite{sun2024evidence}, Cu$_{1-x}$Mn$_{1+y}$SiTe$_3$ \cite{de2025discovery}, and Cu$_2$Cl$_2$ \cite{yang2025multiferroicity}. In contrast, transition-metal dichalcogenides (TMDs) are particularly attractive owing to their rich electronic phase diagram, encompassing charge-density waves, superconductivity, and topological states \cite{manzeli20172d,kolobov2016two,kuc2015electronic}. Their weak van der Waals interlayer bonding enables exfoliation to the monolayer limit while preserving crystalline order, allowing controlled tuning of spin–charge–lattice coupling. Depending on composition, thickness, and structural phase, TMDs exhibit ferroelectricity \cite{sharma2019room} and magnetic ordering \cite{han2013controlling,yun2022escalating}; to date, however, multiferroicity has been experimentally demonstrated only in Te-doped WSe$_2$ \cite{cardenas2023room}. The coexistence of multiple polytypes—2H, 1T', and 1T$_d$—with distinct space-group symmetries makes this family especially suitable for investigating how inversion symmetry, electronic topology, and ferroic order are coupled.

Defects and chemical substitutions strongly influence the electronic ground state of transition-metal dichalcogenides (TMDs) by introducing localized states and lowering crystalline symmetry, thereby creating opportunities for emergent magnetism. In 2H-MoS$_2$, sulfur vacancies generate mid-gap states, promote 1T domain formation within the 2H matrix, and induce a room-temperature ferromagnetic moment of $\sim 0.25 \mu_B$ per Mo atom \cite{cai2015vacancy}. First-principles calculations and spectroscopy show that these vacancies stabilize localized magnetic moments through exchange splitting of defect states arising from local inversion-symmetry breaking \cite{tao2014strain, zhang2020extended}. These effects extend to other molybdenum dichalcogenides: experiments and DFT calculations reveal long-range magnetic order in 2H-MoTe$_2$ (40 K) and 2H-MoSe$_2$ (100 K), originating from randomly distributed defects—mainly metal vacancies and chalcogen–metal antisites—with predicted magnetic moments of 0.9–2.8 $\mu_B$ per defect \cite{guguchia2018magnetism}. In ultrathin PtSe$_2$, Pt vacancies induce local moments on neighboring Se atoms that couple via interlayer Ruderman–Kittel–Kasuya–Yoshida interactions, while carrier injection and optical excitation modulate magnetization in V-doped WS$_2$ and WSe$_2$ \cite{avsar2019defect, ortiz2024transition}.

However, defect-induced magnetism is not guaranteed. Structural relaxation can rehybridize defect orbitals and eliminate partially occupied states, so only sufficiently localized and partially filled states sustain exchange interactions \cite{Haldar2015}. This highlights that both the type and local environment of defects are critical in determining magnetic behavior. One strategy to control these effects is defect engineering through compositional tuning. Electron irradiation experiments in monolayer MoS$_2$ show that sulfur vacancies can be produced, diffuse, and agglomerate into extended line defects whose electronic properties are predicted to be tunable by selectively filling the missing sites with other atomic species \cite{komsa2013point}. These observations demonstrate that combining defect engineering with compositional control provides a versatile route to modulate electronic and magnetic states.

Beyond magnetism, symmetry breaking induced by structural distortions and defects can also stabilize ferroelectric order in TMDs. First-principles studies revealed that a structural instability associated with the metal–semiconductor transition in 1T-MoS$_2$ can drive spontaneous polar distortions, establishing that certain distorted TMD phases lie close to a ferroelectric instability \cite{shirodkar2014emergence}. Subsequent experiments on nonstoichiometric 2D TMDs demonstrated that vacancy-induced symmetry lowering produces robust and spatially extendable piezoelectric and ferroelectric responses \cite{hu2023extendable}. Complementary calculations further indicate that surface vacancies can simultaneously induce local magnetic moments and electric dipoles through charge redistribution and lattice relaxation \cite{li2022modulated}. These findings show that defect-mediated symmetry breaking can stabilize cooperative polar distortions, linking local electronic reconstruction to macroscopic ferroelectric order.

Compositional engineering provides an additional pathway to induce and control ferroelectricity. Machine-learning–assisted first-principles studies predict that targeted elemental substitution can stabilize ferromagnetic–ferroelectric multiferroic states by tuning lattice parameters, orbital hybridization, and electronic instabilities \cite{huang2025machine}. More broadly, experimental observations of interfacial ferroelectricity in rhombohedral-stacked bilayer TMDs demonstrate that subtle symmetry modifications—arising from stacking, defects, or chemical substitution—can generate switchable polarization in layered systems \cite{wang2022interfacial}. These results underscore that compositional and structural perturbations systematically tune the balance between centrosymmetric and polar phases.

Substitutional doping also directly modulates magnetic order. First-principles calculations in monolayers of MoS$_2$, MoSe$_2$, MoTe$_2$, WS$_2$, and WSe$_2$ show that doping with period-four transition metals (Ti, V, Cr, Mn, Fe, Co, or Ni) can induce nonmagnetic, ferromagnetic, or antiferromagnetic states, with Curie and Kosterlitz–Thouless transition temperatures tunable up to 300 K depending on the dopant species \cite{tiwari2021magnetic}. These results demonstrate that targeted substitutional doping systematically enhances magnetic ordering, complementing vacancy engineering and compositional tuning strategies.

In the WSe$_2$–WTe$_2$ system, isovalent chalcogen substitution provides a further symmetry-control parameter. Replacing Se with larger, more polarizable Te expands the lattice, enhances spin–orbit coupling, and modifies inversion symmetry \cite{haldar2015systematic, mentzen1976preparation}. This tuning alters the competition between centrosymmetric and polar configurations and reshapes the electronic structure near the Fermi level \cite{man2023ferroic, yun2020ferromagnetic}. Because Se and Te differ in electronegativity and orbital hybridization, alloying modifies band dispersion and inversion-symmetry breaking, providing a compositional route to engineer ferroic states. Experiments showing room-temperature multiferroicity in Te-doped WSe$_2$ single crystals further indicate that coupled compositional tuning and defect-mediated symmetry breaking can stabilize coexisting magnetic and polar order \cite{cardenas2023room}.

Motivated by these symmetry-engineering strategies, we synthesize single crystals of nonstoichiometric solid solutions of WSe$_2$ and WTe$_2$, i.e., W[Te$_x$Se$_{1-x}$]$_{2(1-\delta)}$, via chemical vapor transport to determine how chalcogen substitution and defects concentration jointly control structural symmetry and ferroic order. X-ray diffraction and Raman spectroscopy reveal systematic lattice distortions and symmetry lowering across the compositional series, while piezoresponse and magnetometry measurements demonstrate the emergence of multiple coexisting ferroic responses at room-temperature. By correlating the Te fraction ($x$) and defect concentration ($\delta$), we construct a configurational phase diagram linking structural distortions, ferroelectric order, and magnetic order in the W–Se–Te system. These results identify defect-mediated symmetry breaking as a central mechanism governing ferroic stabilization in layered dichalcogenides and establish a symmetry-based framework for engineering multiferroicity through coupled composition–defect control.
\section{Materials and Methods}

\subsection{Synthesis of single crystals}\label{synthesis_of_single_crystals}

\subsection{X-ray fluorescence (XRF)}\label{Xray_fluorescence}

Elemental compositions were determined by X-ray fluorescence (XRF) using a PANalytical Epsilon 3XLE spectrometer. As-grown single crystals from each batch were selected to minimize contamination from Te-rich segregates or residual precursor material.

Measured mass percentages were converted to molar ratios $n_\text{Te}/n_\text{W}$ and $n_\text{Se}/n_\text{W}$, from which the Te substitution level $x$ and chalcogen vacancy fraction $\delta$ were extracted assuming the nominal formula W(Te$_x$Se$_{1-x}$)$_{2(1-\delta)}$. 

\subsection{X-ray diffraction (XRD)}\label{X_rays_diffraction}

The crystal structure was examined by X-ray diffraction using a PANalytical Empyrean diffractometer with Cu-$K_\alpha$ radiation. Diffraction patterns were collected in an Eulerian cradle geometry over a $2\theta$ range of $5^\circ$–$100^\circ$.

Crystals from the same batches characterized by XRF were mounted on a silicon zero-background holder. The patterns were dominated by $\{00\ell\}$ reflections, consistent with preferential alignment of the $ab$ planes parallel to the substrate, while additional reflections confirmed bulk crystallinity.

\subsection{Raman spectroscopy}\label{Raman_Spectroscopy}

Raman spectroscopy was performed to assess structural symmetry and lattice dynamics. Spectra were acquired for all compositions using a HORIBA XPLORA spectrometer with 532 nm excitation over the range 50–950~cm$^{-1}$. Fresh surfaces were obtained by mechanical cleavage prior to measurement.

\subsection{Magnetic response}\label{Magnetic_response}

Room-temperature magnetic properties were measured using a vibrating-sample magnetometer (VSM, Lake Shore 7400 Series). Bulk measurements were performed on as-grown single crystals (10–25 mg) from multiple batches. A diamagnetic background, arising primarily from the sample holder and minor ambient degradation, was subtracted from all datasets. This correction does not affect the qualitative magnetic trends discussed in Sec.~\ref{Result_Magnetism}.

\subsection{Piezoelectric and ferroelectric response}\label{Piezo_ferro_response}

Piezoelectric and ferroelectric responses were probed by piezoresponse force microscopy (PFM) at room temperature using an environmental AFM system (Cypher ES, Asylum Research). Measurements were performed under an inert N$_2$/Ar atmosphere ($\sim$200 mbar) to minimize surface degradation and electrostatic contributions.

Dual AC resonance tracking (DART-PFM) enhanced sensitivity near the contact resonance, while switching spectroscopy PFM (SS-PFM) provided local electromechanical hysteresis loops and switching parameters, including the coercive voltage.

Piezoresponse amplitudes are reported in arbitrary units and, when specified, converted to picometers using the simple harmonic oscillator (SHO) model in the low-damping regime. Calibration was performed against a periodically poled lithium niobate (PPLN) reference sample. 
\section{Results and discussion}

We begin by establishing the compositional phase space of W(Te$_x$Se$_{1-x}$)$_{2(1-\delta)}$ and its structural evolution with Te substitution and chalcogen vacancy concentration. X-ray diffraction and Raman spectroscopy are then used to analyze symmetry changes and lattice distortions across the series. Finally, we examine how these structural modifications correlate with the magnetic and piezoelectric responses.
\subsection{Composition, defects, and parameterization}
\label{WSeTe_phase}

The investigated crystals belong to the solid-solution series formed between 2H-WSe$_2$ and 1T$_d$-WTe$_2$. Elemental compositions were determined by X-ray fluorescence (XRF). Since XRF yields bulk-averaged elemental ratios, compositions are expressed as chalcogen-to-metal ratios $(n_{Se}/n_W, n_{Te}/n_W)$. 

\begin{figure*}[htb!]
\centering
\includegraphics[width=1\textwidth]{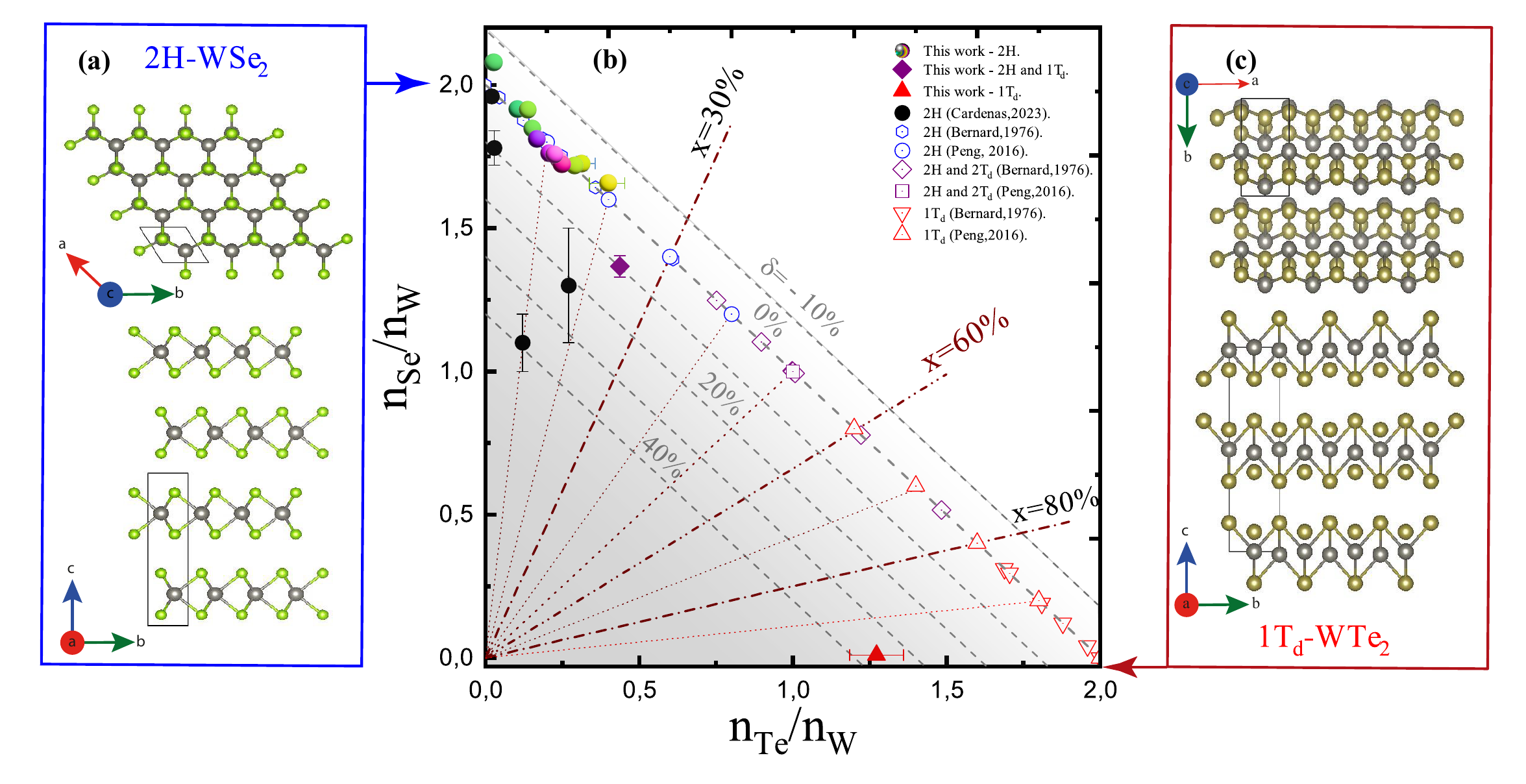}
\caption{
\textbf{(a)} Crystal structure of a 2H-polytype of Te-doped \WSe with space group P63/mmc ($\#$ 194). \textbf{(b)} Diagram of a configurational space representing the possible room-temperature solid solutions with formula \WSeTed. The $x$- and $y$-axes represent the number of moles of tellurium and selenium relative to tungsten, $n_{Te}/n_{W}$ and $n_{Se}/n_{W}$, respectively. The pure components 2H-\WSe and 1T$_d$-WTe$_2$ (with zero defects, $\delta=0$) are represented by the $(0,2)$ and $(2,0)$ coordinates in the diagram. \textbf{(c)} Crystal structure of a 1T$_d$-polytype of \WTe with space group Pnm21. 
}
\label{Phase_Space}
\end{figure*}

To describe both substitution and deviations from ideal stoichiometry within a unified framework, we parameterize the alloys as
\begin{equation}
\text{W}[\text{Te}_x\text{Se}_{1-x}]_{2(1-\delta)},
\end{equation}
where \[
x = \frac{n_{Te}}{n_{Te}+n_{Se}}
\]
defines the Te fraction within the chalcogen sublattice, and $\delta$ quantifies deviations from the ideal MX$_2$ stoichiometry. The total chalcogen-to-tungsten ratio is therefore
\begin{equation}
\frac{n_{Te}+n_{Se}}{n_W} = 2(1-\delta).
\end{equation}
This parameterization separates substitutional disorder caused by Te-doping ($x$) from stoichiometric deviation ($\delta$), establishing them as independent compositional variables.

Figure~\ref{Phase_Space}(b) represents the accessible compositions in the $(n_{Te}/n_W, n_{Se}/n_W)$ plane. Stoichiometric solid solutions ($\delta=0$) lie along the line
\[
\frac{n_{Te}}{n_W} + \frac{n_{Se}}{n_W} = 2,
\]
which connects 2H-WSe$_2$ and 1T$_d$-WTe$_2$. Compositions located away from this line correspond to nonstoichiometric crystals ($\delta \neq 0$), with vertical displacement directly reflecting deviations in total chalcogen content (see Appendix \ref{App_compositional_phase_diagram} for a detailed discussion of the configurational space diagram). Within this framework, $x$ modifies the relative occupation of Se and Te sites, whereas $\delta$ controls the total chalcogen concentration and is therefore directly linked to intrinsic defect populations.

To understand the impact of $\delta$ on the likely defects, defect formation energies reported in previous first-principles studies have been analyzed. Across WSe$_2$ and WTe$_2$, chalcogen vacancies consistently exhibit the lowest formation energies over a broad range of chemical potential conditions, making them the most probable intrinsic defects. Interstitial-related chalcogen configurations generally have higher formation energies under both chalcogen-rich and metal-rich conditions, but are still energetically more favorable than all metal-related defects. In turn, metal vacancies, metal interstitials, chalcogen--metal antisites, and multi-vacancy complexes (e.g., double or higher-order vacancies) form the least favorable defect types in these materials (see Appendix \ref{App_Defec_Hierarchy} for a detailed discussion of defect classification and energy hierarchy). Accordingly, $\delta>0$ corresponds to chalcogen-deficient regimes likely dominated by chalcogen vacancies, whereas $\delta<0$ reflects chalcogen-rich conditions in which chalcogen interstitials the most dominant type of defect.

Despite this theoretical understanding of defect energetics, experimental studies have largely focused on the effects of Te substitution. Stoichiometric \WSeTe\ alloys have been examined in terms of structural evolution \cite{mentzen1976preparation}, electronic transport \cite{yu2017metal,kanchanavatee2022phase}, valleytronic phenomena \cite{oliver2020valley}, epitaxial growth \cite{barton2019wse}, and thermal properties \cite{qian2018anisotropic}. Deviations from chalcogen stoichiometry, by contrast, have received comparatively limited attention. The open symbols in Fig.~\ref{Phase_Space}(b) indicate compositions previously explored in the literature.

To address this underexplored region, the present work focuses primarily on low tellurium content in the range ($x < 25\%$) and deviations from chalcogen stoichiometry in the range ($|\delta| < 20\%$), providing a well-defined compositional window to examine how substitution and stoichiometric variations influence the crystal structure. Because Te substitution ($x$) is expected to dominate structural modifications, while deviations from chalcogen stoichiometry ($\delta$) can subtly tune local bonding environments, we employ X-ray diffraction to probe the structural consequences of Te incorporation across these regimes of defect populations.
\subsection{X-ray diffraction results}\label{Results_DRX}

\begin{figure*}[htb!]
\centering
\includegraphics[width=1\textwidth]{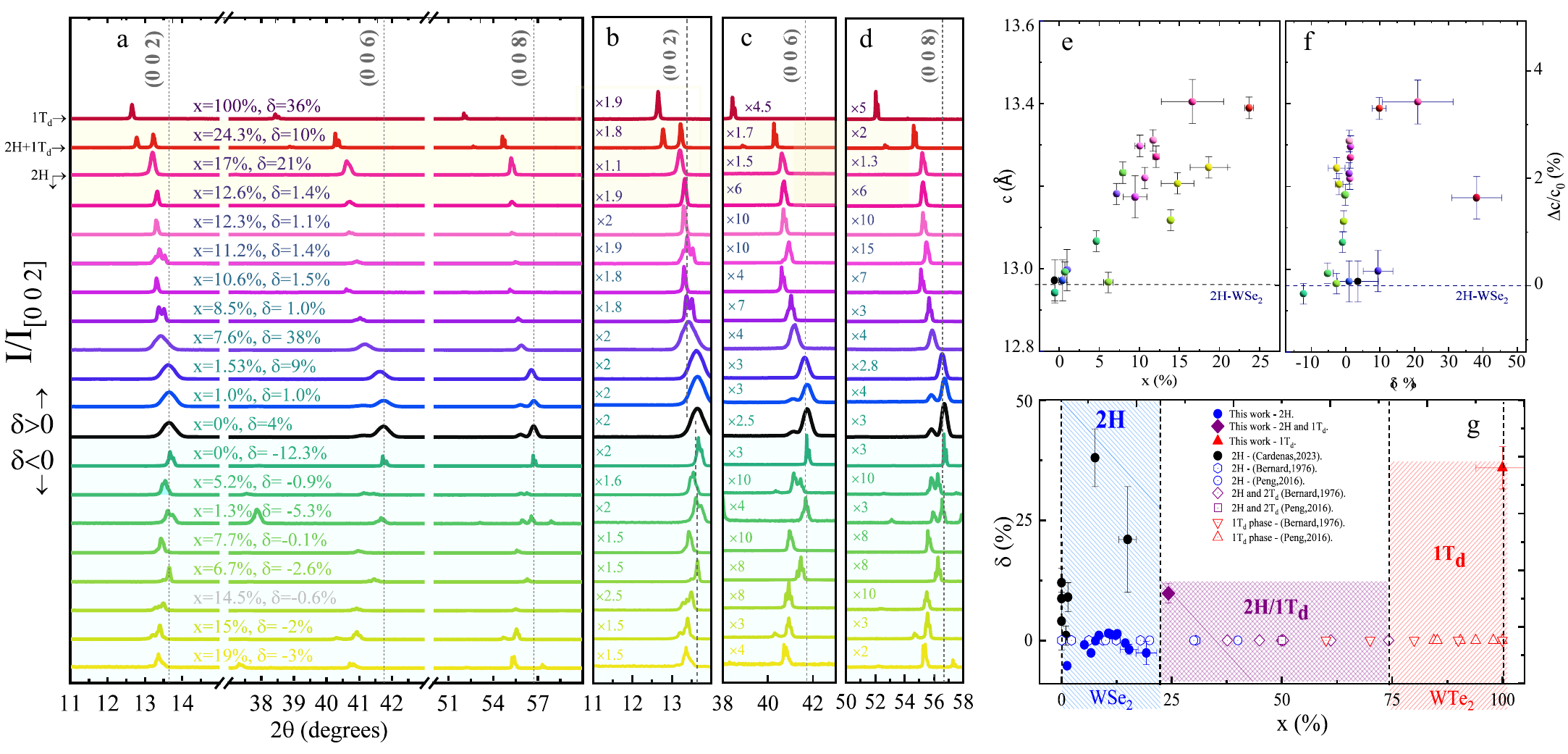}
\caption{\textbf{(a)} Powder x-ray diffraction patterns of undoped and Te-doped samples. The observed peaks correspond to the ${00\ell}$ reflections of the 2H polytype and align with the reported positions of stoichiometric 2H-WSe$_2$ (dotted vertical lines) for compositions with $x \le 17\%$, spanning both $\delta>0$ and $\delta<0$ regimes. \textbf{(b)}–\textbf{(d)} Expanded views of representative reflections [(002), (006), and (008)] showing a systematic shift toward lower $2\theta$ with increasing Te content, consistent with an expansion of the $c$-axis lattice parameter. In the close-up views, patterns are vertically offset and independently scaled in intensity for clarity. \textbf{(e)} Evolution of the extracted $c$-axis lattice parameter as a function of Te content. \textbf{(f)} Dependence of the lattice parameter on the chalcogen defect fraction $\delta$. \textbf{(g)} Schematic representation of structural symmetry across the compositional $x$–$\delta$ phase space.}
\label{DRX}
\end{figure*}

To capture the structural effects of compositional variations, X-ray diffraction (XRD) measurements were performed on the \WSeTed\ crystals. The main results are summarized in Figure~\ref{DRX}. Owing to the flake-like morphology of the crystals, the samples preferentially align along the $a$–$b$ plane during measurements in Eulerian cradle geometry. All specimens were analyzed in bulk form without prior grinding, which enhances the intensity of reflections from the ${00\ell}$ family while suppressing contributions from other crystallographic orientations. As a result, the in-plane lattice parameters ($a$ and $b$) cannot be reliably extracted, whereas the $c$-axis lattice parameter can be determined with high accuracy from the diffraction patterns.

Figure~\ref{DRX} (a–c) displays the evolution of the diffraction patterns as a function of tellurium content ($x$) and chalcogen defect fraction ($\delta$). For clarity, chalcogen-deficient compositions ($\delta > 0$) are shown at the top of the panels, while chalcogen-rich ones ($\delta < 0$) appear at the bottom. Within each regime, the Te fraction increases upward for $\delta > 0$ and downward for $\delta < 0$, enabling a direct visualization of the compositional trends across the series.

Dashed vertical lines in Figure~\ref{DRX}(a) indicate the principal diffraction peaks of the 2H polytype of pure 2H-WSe$_2$, which serve as reference benchmarks. Undoped samples ($x = 0$), measured under both chalcogen-deficient ($\delta = 4\%$) and chalcogen-rich ($\delta = -12.3\%$) conditions, exhibit peak positions that closely match reported values for 2H-WSe$_2$ across the explored range of non-stoichiometry. In contrast, systematic peak shifts become evident as the Te fraction increases, independent of whether the samples are chalcogen-deficient or chalcogen-rich.

Close-up views of representative ${00\ell}$ reflections are presented in Figures~\ref{DRX} (b–d). These panels reveal a progressive displacement of the peaks toward lower diffraction angles with increasing Te concentration, consistent with an expansion of the unit cell. This behavior is expected from the larger atomic radius of Te relative to Se ($r_\text{Te} = 1.18\,r_\text{Se}$; see Appendix~\ref{Atomic_Radii}).

The compositional dependence of the lattice parameter is quantified in Figure~\ref{DRX} (e), which shows that the $c$-axis increases monotonically with Te incorporation up to a critical composition $x_c \approx 17\%$. In contrast, no clear systematic correlation is observed between the $c$-axis parameter and the chalcogen defect fraction within $-7\% < \delta < 38\%$ (Figure~\ref{DRX}(f)). While some samples exhibit minor variations (up to $\sim 3\%$), these changes lack a consistent trend, indicating that Te substitution dominates the lattice expansion.

Within the low-doping regime ($x \lesssim x_c$), the $c$-axis lattice parameter follows a linear dependence on composition (consistent with the close-packing model described in Appendix~\ref{App_hcp}), $c(x,\delta) = m(\delta)x + b(\delta)$, where $m(\delta)$ and $b(\delta)$ represent composition-dependent coefficients corresponding to the incremental expansion per unit Te content and the baseline lattice spacing for a given chalcogen defect fraction, respectively.

Beyond the systematic peak shifts, qualitative differences in peak shape are observed depending on $\delta$. In particular, certain reflections—such as (002) and (008)—exhibit a double-peak structure under chalcogen-rich conditions ($\delta < 0$), whereas broader single peaks are typically observed under chalcogen-deficient conditions ($\delta > 0$). The appearance of double-peak features under chalcogen-rich conditions suggests the coexistence of regions with slightly different interlayer spacings. Such behavior may arise from local structural heterogeneity associated with interstitial chalcogen incorporation, which can generate strong localized strain fields and possibly lead to spatial variations in lattice parameters. In contrast, randomly distributed vacancies are more likely to produce microstrain and peak broadening rather than well-resolved peak splitting, as they modify the average lattice parameter more uniformly in the absence of defect clustering.

Importantly, within the low-Te regime ($x < x_c$), no additional reflections associated with alternative crystal structures are detected. The persistence of shifted 2H-WSe$_2$-like peaks and the absence of extra diffraction features indicate that both Te substitution and chalcogen non-stoichiometry preserve the global hexagonal symmetry of the host lattice.

However, for $x > x_c$, the solid-solution limit of the 2H structure is exceeded. As shown in Figure~\ref{DRX}, additional reflections emerge that cannot be indexed within the hexagonal symmetry, signaling the onset of a structural phase transition toward the orthorhombic 1T$_d$ phase characteristic of WTe$_2$. The $2H \rightarrow 1T_d$ transition corresponds to a change in the tungsten coordination environment from trigonal prismatic to distorted octahedral. This transformation is driven by Te substitution, which modifies the local bonding environment due to its larger atomic radius and lower electronegativity ($\chi_\text{Te} = 2.1$ vs.\ $\chi_\text{Se} = 2.4$). Near the transition boundary, coexistence of 2H and 1T$_d$ reflections is observed, consistent with the decreasing energy difference between these polytypes as Te content increases~\cite{oliver2020valley}. At still higher Te concentrations, the structure evolves toward the fully orthorhombic 1T$_d$–WTe$_2$ phase as $x \rightarrow 1$~\cite{mentzen1976preparation}.

Overall, the proportional peak shifts with Te incorporation reflect a systematic increase in lattice parameters and unit-cell volume. Even for chalcogen deficiencies as large as $\delta = 38\%$, no vacancy-induced phase transitions are detected, underscoring the structural robustness of the 2H-WSe$_2$ framework within the explored compositional window.
While X-ray diffraction probes the average crystallographic structure, it is less sensitive to local bonding modifications and short-range disorder. Because Te substitution and chalcogen-related defects can perturb lattice vibrations and local symmetry, Raman spectroscopy was employed as a complementary probe.
\subsection{Raman spectroscopy}
\label{Results_Raman}

\begin{figure*}[htb!]
\centering
\includegraphics[width=1\textwidth]{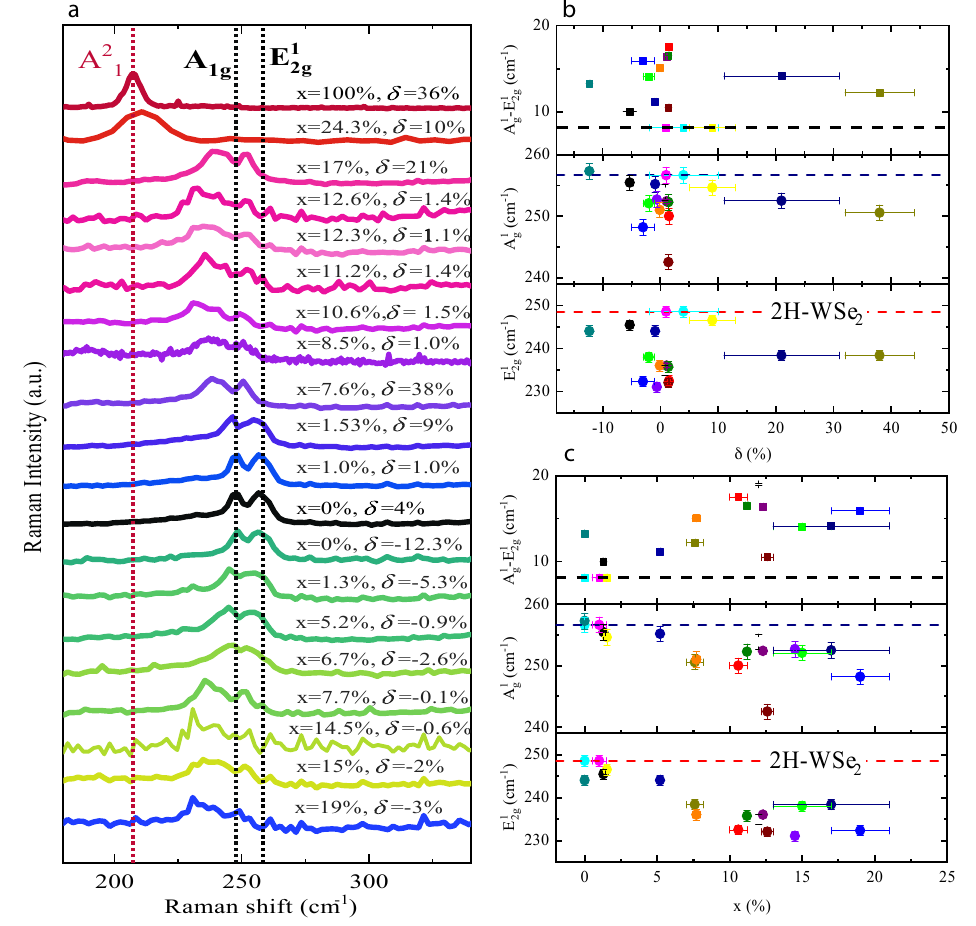}
\caption{\textbf{(a)} Raman spectra of 
W[Te$_x$Se$_{1-x}$]$_{2(1-\delta)}$ single crystals with varying composition showing a systematic shift toward lower wavenumbers with increasing Te content. Dashed lines mark the peak positions of the undoped ($x=0$) compound. \textbf{(b)} Evolution of the extracted $E^1_{2g}$ and $A^1_g$ peak positions, their separation $(A^1_g-E^1_{2g})$, and relative intensities as functions of Te content. \textbf{(c)} Corresponding evolution as a function of the chalcogen defect fraction $\delta$.}
\label{Raman}
\end{figure*}

Raman spectra collected for samples with different stoichiometries are shown in Fig.~\ref{Raman}(a). The spectrum of stoichiometric 2H-WSe$_2$ is displayed as a reference (black curve), with the characteristic $E^1_{2g}$ and $A^1_g$ vibrational modes indicated by dashed vertical lines. All doped samples retain the defining features of the 2H-WSe$_2$ phase, corroborating the structural stability inferred from X-ray diffraction.
 
Figure~\ref{Raman}(a) shows that both the $E^1_{2g}$ and $A^1_g$ modes persist across the compositional series. The $E^1_{2g}$ mode corresponds to in-plane atomic vibrations within a single layer, whereas the $A^1_g$ mode involves out-of-plane vibrations of chalcogen atoms in adjacent layers. Both modes shift systematically toward lower wavenumbers with increasing Te content, with the redshift of $E^1_{2g}$ being more pronounced.

A useful framework for interpreting these trends is to consider each vibrational mode as a harmonic oscillator with frequency, $\omega \propto \sqrt{\frac{k(x,\delta)}{\mu(x,\delta)}}$, where $\mu(x,\delta)$ is the effective mass and $k(x,\delta)$ represents the effective bonding stiffness. Te substitution increases the average chalcogen mass and modifies local bonding environments, both of which influence the phonon frequencies.

The absence of additional Raman-active features, together with the systematic redshift, indicates that Te incorporation primarily affects the vibrational spectrum through mass and lattice-expansion effects rather than through symmetry breaking or secondary phase formation. Figures~\ref{Raman}(b)–(c) summarize the evolution of the $E^1_{2g}$ and $A^1_g$ peak positions, as well as their separation $(A^1_g -E^1_{2g})$, as functions of Te content and chalcogen defect fraction. The shifts correlate strongly with the Te concentration $x$ and show comparatively weaker dependence on $\delta$, consistent with a dominant substitution-driven effect.

Changes in bond length associated with lattice expansion, as revealed by X-ray diffraction, likely contribute to the observed frequency reduction through a softening of effective force constants. The combined XRD and Raman results therefore provide a consistent picture of gradual lattice expansion and preserved 2H symmetry across the investigated compositional range.

While Raman spectroscopy probes local bonding and lattice dynamical modifications, magnetic measurements provide insight into how compositional disorder and defect populations influence spin interactions. In particular, variations in Te substitution ($x$) and chalcogen stoichiometry ($\delta$) allow us to examine the relationship between structural perturbations and magnetic response in these layered dichalcogenides.

\subsection{Magnetic Properties}\label{Result_Magnetism}

The magnetic properties of the \WSeTed\ alloys were investigated as a function of tellurium fraction $x$ and chalcogen defect parameter $\delta$. Magnetization curves $M(H)$ measured at 300 K for representative compositions are shown in Figs.~\ref{Magnetism}(a,b), after subtraction of diamagnetic contributions.

\subsubsection*{Defect-induced magnetism}


Nearly stoichiometric samples ($|\delta| < 5\%$) exhibit an $M(H)$ behavior without measurable hysteresis, consistent with weak paramagnetism across the explored Te range. In contrast, chalcogen-deficient compositions (positive $\delta$) display progressively open hysteresis loops, indicative of ferromagnetic ordering, particularly for $\delta > 10\%$.

These observations suggest a gradual evolution from paramagnetic to ferromagnetic behavior as defect density increases, pointing toward a defect-mediated origin of the magnetic moments. Prior theoretical and experimental studies in transition-metal dichalcogenides have shown that chalcogen vacancies can induce localized magnetic moments by altering the local electronic structure and partially depopulating $p$–$d$ hybridized states~\cite{cardenas2023room,Ataca2012stable,Yazyev2010emergence,Tongay2012magnetic,Zhao2019recent,Shidpour2010density}, consistent with the compositional dependence observed here.

\begin{figure*}[htb!]
\centering
\includegraphics[width=1\textwidth]{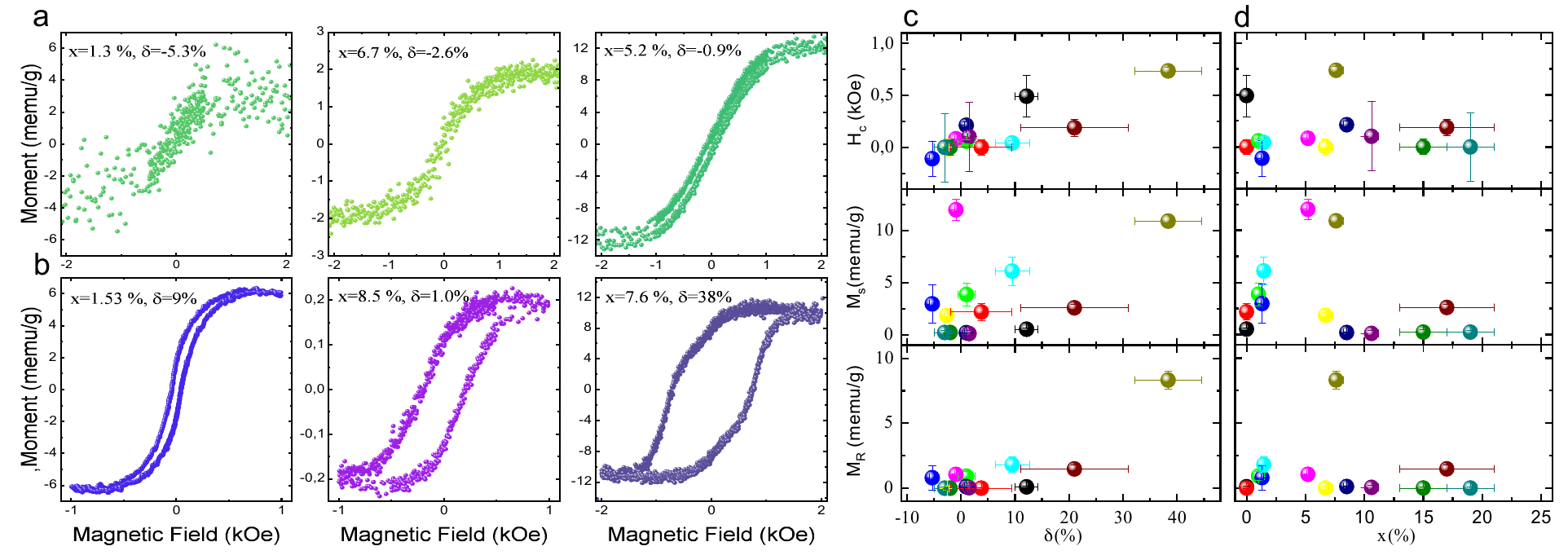}
\caption{Magnetization hysteresis loops at 300 K for \WSeTed\ samples with different tellurium fractions $x$, for both (a) $\delta < 0$ and (b) $\delta > 0$. Diamagnetic contributions have been subtracted. Panels (c) and (d) display the coercive field $H_c$, saturation magnetization $M_s$, and remanent magnetization $M_R$ as functions of the chalcogen defect parameter $\delta$ and tellurium fraction $x$, respectively.}
\label{Magnetism}
\end{figure*}

To quantify these effects, we extracted the coercive field ($H_c$), remanent magnetization ($M_R$), and saturation magnetization ($M_s$). Figure~\ref{Magnetism}(c) shows that both $H_c$ and $M_R$ increase systematically with $\delta$, whereas Fig.~\ref{Magnetism}(d) shows a more complex dependence on Te fraction. The saturation magnetization predominantly increases with vacancy concentration and shows a complex, nonmonotonic dependence on Te substitution at fixed $\delta$, reflecting the interplay between defect connectivity and doping, as suggested by the experimental observations.

Although the present measurements do not directly reveal the exchange mechanism, the systematic enhancement of magnetism with increasing $\delta$ suggests stronger interactions among defect-induced moments. Chalcogen vacancies create electronic states near the Fermi level that can hybridize with host states and provide itinerant carriers mediating indirect exchange, resembling Ruderman–Kittel–Kasuya–Yosida (RKKY) coupling. In three-dimensional metals, this exchange oscillates with the Fermi wave vector and decays as $J(r)\propto r^{-3}$~\cite{ruderman1954indirect,kasuya1956theory,yosida1957magnetic}, whereas in quasi-two-dimensional systems the decay is slower and can exhibit enhanced anisotropy~\cite{mastrogiuseppe2014rkky,Zhou2010strength}. While RKKY-like coupling is plausible in our alloys, it should be regarded as a guiding framework rather than a conclusive mechanism. In two-dimensional semiconductors, randomly distributed magnetic ions can generate long-range ferromagnetic order via RKKY~\cite{stephanovich2021carrier}, and theoretical studies in semiconducting transition-metal dichalcogenides predict that localized moments can likely sustain tunable long-range order through edge deposition of magnetic impurities, midgap-state selection via doping~\cite{avalos2018long,parhizgar2013indirect}, and electric-field control of RKKY interactions~\cite{mousavi2021electrical}.

In this sense, a plausible explanation for the enhancement of magnetic response with defect density can be found in the following argument: At low vacancy densities ($\delta \lesssim 5\%$), moments remain largely isolated, producing a paramagnetic response. As $\delta$ increases, the reduction in inter-defect spacing enhances exchange interactions, allowing spin clusters to form and eventually percolate into long-range ferromagnetic order~\cite{Tongay2012magnetic,Ataca2012stable,cai2015vacancy}, consistent with the measured paramagnetic-to-ferromagnetic transition.

\subsubsection*{Tellurium-Induced Modulation of Defect Magnetism}

The influence of Te substitution on magnetic behavior is reflected by the experimental data. The magnetic response as a function of Te fraction $x$ [Fig.~\ref{Magnetism}(d)] exhibits a nonmonotonic behavior: the saturation magnetization $M_s$ increases at low $x$, reaches a maximum at intermediate compositions, and decreases at higher $x$. This behavior suggests that Te affects both the connectivity of the defect network and the effective magnetic interactions among vacancy-induced moments.

At low $x$, the increase in $M_s$ reflects probable enhanced stabilization of defect moments, while the maximum at intermediate compositions likely corresponds to near-optimal connectivity of the defect network. To describe this conceptually, we introduce the effective magnetic coupling $J^\mathrm{eff}_\mathrm{M}(x,\delta)$, defined as the average exchange interaction among defect-induced magnetic moments, rather than a directly measured quantity, as a function of Te fraction $x$ and chalcogen defect parameter $\delta$. Near-optimal connectivity at intermediate $x$ may enhance $J^\mathrm{eff}_\mathrm{M}(x,\delta)$, contributing to the observed peak in $M_s$. At higher Te content, the decrease in $M_s$ coincides with the approach to the semimetallic 1T$_d$-WTe$_2$ phase~\cite{ali2014large,soluyanov2015type}, where increased carrier itinerancy and possible spin-flip scattering via the Elliott–Yafet mechanism~\cite{Elliott1954,Yafet1963,kurpas2021intrinsic} can partially reduce the stability of defect-localized moments.

The underlying mechanism can be understood in terms of spin–orbit coupling (SOC). In \WSeTed, chalcogen vacancies create localized defect states with significant W 5$d$ character due to disrupted W–$X$ hybridization~\cite{komsa2012two}, which can host magnetic moments. In a quasi-two-dimensional lattice, the stability of long-range magnetic order is strongly influenced by spin anisotropy. According to the Mermin–Wagner theorem~\cite{mermin1966absence,gibertini2019magnetic}, finite-temperature ordering requires breaking of continuous spin-rotational symmetry. 
Te substitution introduces symmetry breaking by enhancing SOC. The SOC Hamiltonian, $H_{\mathrm{SOC}} = \lambda_\mathrm{SOC} \mathbf{L} \cdot \mathbf{S}$, scales approximately as the fourth power of the atomic number, $Z^4$, so replacing Se ($Z = 34$) with Te ($Z = 52$) strengthens SOC by a factor of $\left(\frac{52}{34}\right)^4 \approx 5$. Because the vacancy-induced $d$ states are strongly hybridized with chalcogen $p$ orbitals, this enhanced SOC directly affects the defect-localized moments. The enhanced anisotropy increases the stability of spin alignment against thermal fluctuations, thereby promoting ferromagnetic order~\cite{zhou2013intrinsic,riley2014direct,gong2019two}.

Overall, magnetic ordering in \WSeTed\ depends primarily on the chalcogen defect parameter $\delta$, while Te substitution serves as a secondary tuning factor that modulates both the connectivity of defect moments and the strength of SOC-induced stabilization.
\subsection{Piezoresponse Force Microscopy (PFM) Characterization}\label{Result_Piezoelectricity}

Having detailed the influence of vacancies and Te substitution on the magnetic behavior, we next examine how the same defects affect local polar properties. In layered W[Te$_x$Se$_{1-x}$]$_2(1-\delta)$, chalcogen vacancies and compositional tuning not only modify the magnetic exchange network but can also locally break inversion symmetry, potentially giving rise to polar regions. To probe this effect, switching-spectroscopy (SS-PFM) and dual AC resonance tracking (DART-PFM) measurements were performed across the compositional series.

\subsubsection*{Butterfly and non-butterfly responses: ferroelectric versus piezo/paraelectric behavior}

The studied samples span Te fractions from $x \sim 1\%$ to $17\%$ and defect regimes from $\delta \sim -3\%$ to $38\%$, allowing us to correlate defect density, Te content, and local piezoelectric response. SS-PFM measurements reveal two distinct classes of electromechanical behavior, which depend systematically on the global stoichiometry and defect content.

In the \textit{ferroelectric regime}, vacancy-rich samples  exhibit amplitude–voltage curves with a symmetric butterfly shape accompanied by a $180^\circ$ phase reversal [Fig.~\ref{Piezo}(a)]. This combination is commonly associated with switchable polarization in PFM measurements~\cite{alikin2022exploring,aggarwal2014direct,miao2014more}, reflecting field-induced reversible reorientation of local dipoles under cyclic bias, although contributions from electrostatic interactions or bias-induced charge trapping cannot be fully excluded in PFM measurements \cite{vasudevan2017ferroelectric}.

Conversely, in the \textit{piezo/paraelectric regime}, observed in low-vacancy samples, the phase signal switches with bias polarity but shows no hysteretic loop (Fig.~\ref{Piezo}a). Here, the response is consistent with linear piezoelectric coupling or field-induced alignment of paraelectric dipoles~\cite{bonnell2009piezoresponse,balke2015differentiating,zhang2021pfm}.

These contrasting behaviors suggest that the emergence of ferroelectric-like response correlates primarily with global composition and defect density, rather than arising solely from local variations within individual crystals.

\subsubsection*{Defect-induced polarization and local piezoelectric response}

The microscopic origin of the ferroelectric behavior is inferred from amplitude and phase maps acquired across multiple locations within each crystal. Vacancy-rich crystals ($\delta > 20\%$) exhibit local hysteresis in both amplitude and phase, consistent with switchable polarization at the nanoscale. In contrast, low-vacancy crystals display only a linear electromechanical response. Moderate negative deviations from stoichiometry ($\delta \approx -1\%$ to $-3\%$) yield measurable piezoresponse, likely arising from asymmetry induced by interstitial defects. Collectively, these observations suggest that the connectivity of defect-induced polar sites may contribute to stabilizing cooperative ferroelectric-like behavior, although direct structural evidence of long-range dipolar order is beyond the scope of the present measurements.

\subsubsection*{Compositional dependence}
Spatially resolved amplitude and phase maps confirm that the observed responses are reproducible across multiple locations within each crystal. Near-stoichiometric samples consistently exhibit linear electromechanical behavior, whereas vacancy-rich compositions show localized hysteretic switching.

First-principles calculations suggest that chalcogens carry positive Hirshfeld charges, while W atoms carry roughly twice the negative charge, producing local dipoles~\cite{cardenas2023room}. Te atoms exhibit larger charges than Se (up to 2–3 times), which may enhance the magnitude of local dipole moments in the presence of lattice distortions. These observations are consistent with a correlation between Te-induced lattice distortions and enhanced local piezoelectric response, although a quantitative link requires further theoretical investigation.

\begin{figure*}[htb!]
    \centering
    \includegraphics[width=1\textwidth]{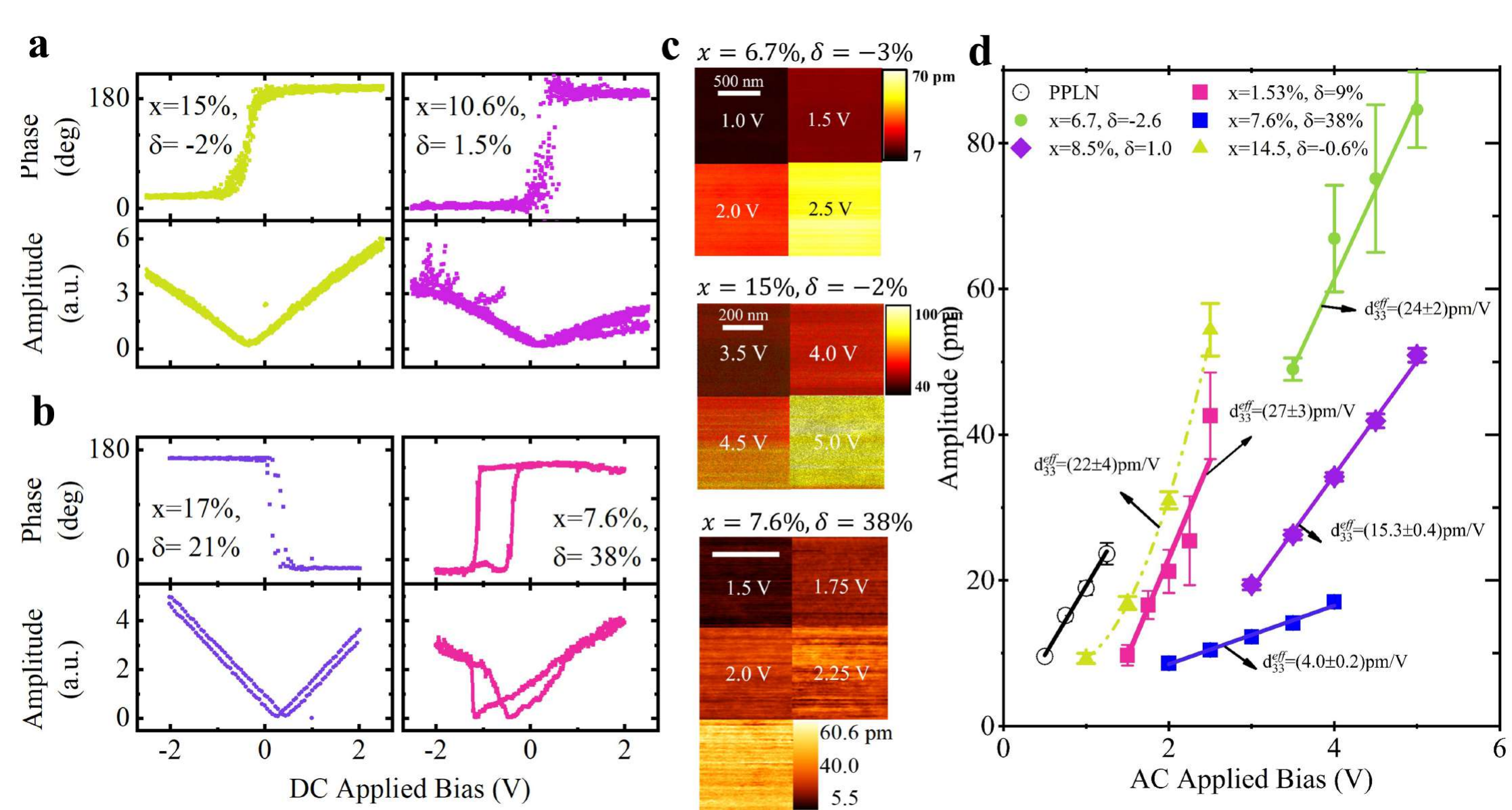}
    \caption{SS-PFM piezoresponse phase and amplitude loops at 300 K for \WSeTed\ samples with varying tellurium fractions $x$. \textbf{(a)} $\delta < 0$ and \textbf{(b)} $\delta > 0$. \textbf{(c)} DART-PFM amplitude normalized by the simple harmonic oscillator (SHO) model for $x=7\%, \delta=-2\%$. \textbf{(d)} Effective $d_{33}$ piezoelectric coefficient from DART-PFM amplitude measurements for $x=7\%, \delta=-2\%$ and $x=9\%, \delta=-1\%$. Error bars correspond to the standard deviation across each $500$~nm measurement area for each voltage.}
    \label{Piezo}
\end{figure*}

\subsubsection*{Effective piezoelectric coefficient}

Nanoscale amplitude DART-PFM maps identify regions with measurable electromechanical response [Fig.~\ref{Piezo} (c)], corresponding to local polar distortions induced by chalcogen vacancies and Te substitution. The consistency of the response across spatially separated regions indicates that the behavior is representative of each compositional batch rather than arising from isolated inhomogeneities.

Across multiple crystals, the effective piezoelectric coefficient $d_{33}^{\rm eff}$ increases with total chalcogen content, from $\sim 15.6$~pm/V at $x \sim 8\%$ to $\sim 22.5$~pm/V at $x \sim 17\%$ [Fig.~\ref{Piezo} (d)]. This trend may be rationalized in terms of lattice symmetry and local structural distortions, as low Te content favors the symmetric 2H phase, whereas higher Te fractions are associated with locally distorted 1T$_d$ regions, potentially enhancing local polar response.

Taken together, the PFM results suggest that ferroelectric-like behavior becomes more prominent above a critical vacancy density, while lower-defect samples remain in a piezoelectric or paraelectric regime. However, complementary structural or macroscopic polarization measurements would be required to unambiguously establish long-range ferroelectric order.

\section{Ferroic Phase Diagram}\label{Ferroic_phase_diagram} 

\begin{figure*}[htb!]
\centering
\includegraphics[width=1\textwidth]{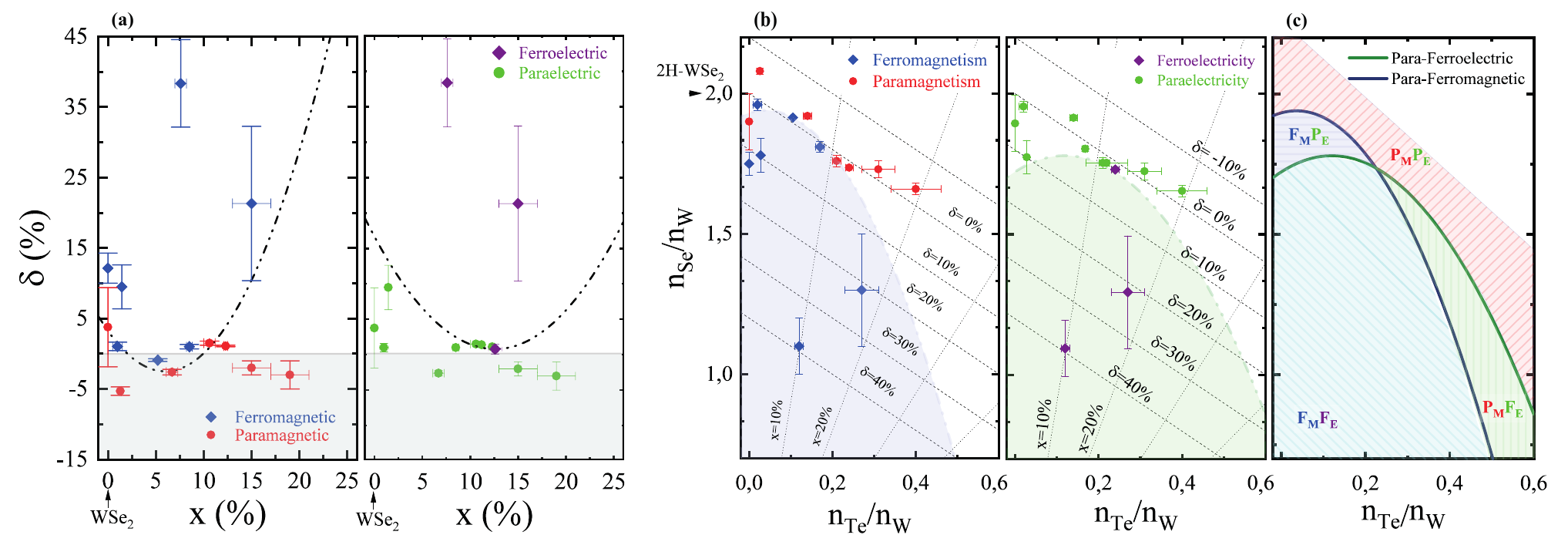}
\caption{Para/ferro-magnetic/electric regions represented in (a) the $x$–$\delta$ diagram and (b) the $n_\text{Te}/n_\text{W}$–$n_\text{Se}/n_\text{W}$ plane. (c) Schematic summary highlighting regions of multiferroicity, paraelectric–ferromagnetic, paramagnetic–ferroelectric-like, and paramagnetic–paraelectric behavior. Dashed boundaries correspond to empirical separatrices between ferroic and para states.}
\label{FerroicDiagram}
\end{figure*}

To investigate ferroic states in W(Se,Te)$_2$, we constructed configurational phase diagrams in compositional spaces defined by $(\frac{n_\text{Te}}{n_\text{W}}, \frac{n_\text{Se}}{n_\text{W}})$ and $(x, \delta)$. Figure~\ref{FerroicDiagram} (a) shows the ferroic phase diagram in $(x,\delta)$, with all samples retaining 2H-WSe$_2$ symmetry, while Fig.~\ref{FerroicDiagram} (b) presents the same trends in absolute compositional ratios. Blue squares denote ferroic order—ferromagnetic or ferroelectric-like —and red circles indicate paraelectric/piezoelectric or paramagnetic. Dashed boundaries follow the empirical form $(\delta - \delta_0) = A (x - x_0)^2$, which approximate composition-dependent critical vacancy fractions. Here, $x_0$ corresponds to the Te fraction minimizing the critical vacancy density $\delta_c$, with $\delta_0 \approx \delta_c(x_0)$.

Ferromagnetic and ferroelectric-like responses emerge at large positive $\delta$, consistent with vacancy-induced local moments and defect-stabilized lattice distortions in transition-metal dichalcogenides~\cite{Tongay2012magnetic,Yazyev2010emergence}. In contrast, negative $\delta$ ($\delta<0$), associated with chalcogen-rich compositions, correlates with paraelectric and paramagnetic behavior, where reduced vacancy density and local symmetry breaking are insufficient to stabilize long-range ordering.

More fundamentally, the separatrices $\delta_c^{(M)}(x)$ and $\delta_c^{(P)}(x)$ mark composition-dependent critical boundaries for the magnetic ($M$) and polar ($P$) order parameters arising from the same defect population. Because magnetic moments and electric dipoles originate from chalcogen vacancies but are mediated by distinct microscopic interactions, their ordering thresholds do not necessarily coincide. This naturally generates ferromagnetic–paraelectric and paramagnetic–ferroelectric-like regimes, as well as multiferroic states when both thresholds are exceeded.

The phase diagram therefore comprises four regimes [Fig.~\ref{FerroicDiagram}(c)]: (i) paramagnetic–paraelectric ($M=0$, $P=0$), (ii) paramagnetic–ferroelectric-like ($M=0$, $P\neq0$), (iii) ferromagnetic–paraelectric ($M\neq0$, $P=0$), and (iv) multiferroic ($M\neq0$, $P\neq0$) with coexistence of nonzero magnetization and polarization. The curved boundaries show that the critical vacancy densities for magnetic and polar ordering, $\delta_c^{(M)}(x)$ and $\delta_c^{(P)}(x)$, depend on composition. Te substitution therefore tunes how effectively vacancies stabilize each type of order, yielding a composition-controlled system with two interacting order parameters.

Although magnetoelectric coupling has not been experimentally determined yet for this system, the coexistence and composition-dependent onset of both magnetic ($M$) and polar ($P$) orders are consistent with a two-order-parameter Landau description~\cite{Landau1937}. In this framework, vacancy density and composition modify the stability conditions for each instability, and because both originate from chalcogen vacancies, the $(x,\delta)$ diagram represents a defect-controlled, composition-tuned phase space.
\section{Discussion}\label{Discussion}

The phase diagram behavior of W(Se,Te)$_2$ can be understood within a hierarchical framework separating structural, electronic, and geometric contributions. In the composition range investigated ($x\leq x_C$), the crystal retains 2H symmetry, and no structural phase transition is detected. The observed ferroic responses therefore originate from defect-induced modifications of the electronic ground state rather than global symmetry breaking. Higher Te concentrations can drive structural transformations, but the phenomena discussed here occur entirely within a single crystallographic phase, governed primarily by chalcogen substitution and chalcogen nonstoichiometry.

\subsection{Defect-Induced Magnetic and Polar States}

In 2H-TMDs, W atoms form hexagonally coordinated layers, defining a triangular lattice in the basal plane \cite{cardenas2023room,freire2022vacancy}. The W magnetic moment is highly sensitive to its local bonding environment; first-principles calculations show strong dependence on the presence, arrangement, and concentration of neighboring chalcogen atoms or vacancies \cite{cardenas2023room,li2016strong}.

Chalcogen vacancies introduce partially occupied states near the Fermi level, which can undergo exchange splitting to stabilize W-centered magnetic moments \cite{hong2015exploring,komsa2012two,cai2015vacancy,cardenas2023room}. Magnetic sites occupy a fraction $p(\delta)$ of lattice sites, and vacancy-rich compositions favor interacting moments via carrier-mediated exchange or superexchange. The observed magnetic hysteresis at large $\delta$ is consistent with defect-activated magnetism above a critical vacancy fraction.

Vacancies also break local inversion symmetry, generating electric dipoles directed from the W plane toward remaining chalcogen atoms \cite{cardenas2023room,yang2018origin,jiang2025two}. Each vacancy therefore contributes both a magnetic moment and a dipole. While the average structure remains centrosymmetric, a sufficient density of locally polar regions enables correlated dipolar switching detectable by PFM.

The coexistence of vacancy-induced magnetic moments and dipoles motivates a site-diluted triangular-lattice model with coupled spin (magnetic) and pseudospin (dipolar) variables. The near-coincidence of magnetic and polar thresholds in the phase diagram suggests a common microscopic origin: symmetry-breaking defects destabilizing the local electronic configuration, simultaneously driving magnetization and polarization.

Although vacancies in TMDs can be dynamic at elevated temperatures~\cite{flototto2025large,song2017probing}, in W(Se$_{1-x}$Te$_x$)$_{2(1-\delta)}$ the combination of Se/Te substitution ($x$) and chalcogen vacancies ($\delta$) effectively freezes the defect configuration during growth~\cite{hossen2024defects,afrid2024defect}. Consequently, for modeling and interpretation of room-temperature behavior, these vacancies can be treated as quenched~\cite{gusakov2022formation,mitterreiter2021role}. This static disorder induces local variations in bond geometry and symmetry-breaking fields, generating spatially inhomogeneous exchange and polar couplings \cite{kirubasankar2022atomic,khalid2024deep}. Alloy and vacancy disorder further modulate structural phase stability, induce mid-gap states, modify Raman modes, and alter other electronic and optical properties \cite{schuler2019large}, naturally promoting clustered or percolative ferroic states with macroscopically non-uniform responses \cite{hoyos2006percolation,deng2004magnetic}.
\subsection{Site-Diluted Coupled Ising Framework}

The site-diluted coupled Ising (SDCI) picture provides a unified framework to describe ferroic behavior in W(Se$_{1-x}$Te$_x$)$_{2(1-\delta)}$. In this model, vacancies act as active lattice sites hosting both spin ($S_i=\pm1$) and dipolar ($P_i=\pm1$) degrees of freedom:
\begin{equation}
\mathcal{H} = \mathcal{H}_{\mathrm{M}} + \mathcal{H}_{\mathrm{E}} + \mathcal{H}_{\mathrm{ME}},
\end{equation}
with
\[
\mathcal{H}_{\mathrm{M}} = -\sum_{\langle i,j\rangle} J^{\rm eff}_M(x,\delta)\,\eta_i \eta_j S_i S_j - h^M \sum_i \eta_i S_i,
\]
\[
\mathcal{H}_{\mathrm{E}} = -\sum_{\langle i,j\rangle} J^{\rm eff}_E(x,\delta)\,\eta_i \eta_j P_i P_j - h^E \sum_i \eta_i P_i,
\]
\[
\mathcal{H}_{\mathrm{ME}} = -\lambda(x,\delta) \sum_i \eta_i S_i P_i,
\]
where $\eta_i \in \{0,1\}$ denotes vacancy-activated sites. Here, $\mathcal{H}_{\mathrm{M}}$ and $\mathcal{H}_{\mathrm{E}}$ describe site-diluted Ising interactions for magnetism and ferroelectricity, respectively, while $\mathcal{H}_{\mathrm{ME}}$ captures spin–polarization coupling arising from spin–orbit and exchange-striction mechanisms \cite{khomskii2009classifying,yahia2017recognition,dong2019magnetoelectricity,zeng2022high}.

The effective magnetic exchange, $J_M^{\rm eff}(x,\delta)$, quantifies the interactions between vacancy-induced W-centered moments. These interactions can arise from multiple mechanisms: for nearby vacancies, superexchange via chalcogen orbitals dominates, while for more distant vacancies, itinerant carriers can mediate long-range Ruderman--Kittel--Kasuya--Yosida (RKKY)-type exchange. Consequently, $J_M^{\rm eff}(x,\delta)$ generally increases with both the vacancy concentration $\delta$ and the efficiency of carrier-mediated coupling. Analogously, $J_E^{\rm eff}(x,\delta)$ describes effective dipolar interactions, which depend on local lattice distortions and the geometric connectivity of polar defect sites.

Within a mean-field approximation, the ordering temperature scales as
\[
k_B T_c \sim z \, J^{\rm eff}(x,\delta) \, p(\delta),
\]
where $p(\delta)$ is the fraction of active sites and $z$ is the effective coordination number \cite{de1993mean}. This relation captures the qualitative trend that $T_c$ increases with both the fraction of active magnetic sites and the effective exchange strength. However, it generally overestimates the actual ordering temperature because it neglects thermal fluctuations and spin correlations in two-dimensional lattices. 

Hysteresis emerges above a critical vacancy fraction, $p_c \simeq \delta_c$, reflecting a percolation-like transition \cite{hoyos2006percolation,deng2004magnetic,zeng2022high}. Spatial variations in local defect density can cause clusters of spins and dipoles, even when co-located, to percolate at different temperatures, naturally explaining the staggered onset of magnetic and polar order in W(Se,Te)$_2$~\cite{cardenas2023room,yang2018origin,afrid2024defect}.

\subsection{Percolative Character of Ferroic Boundaries}

The curvature of the ferroic boundaries reflects the nonlinear dependence of the effective magnetic and polar couplings, $J_M(x,\delta)$ and $J_P(x,\delta)$, on composition. Within a generalized site-diluted Ising/percolation framework~\cite{ballesteros1997ising,stauffer2018introduction}, these couplings are modulated by the fraction of active sites, $p(\delta,x)$, which determines the percolation thresholds for magnetic and electric ordering. 

We describe the system using a coupled double Ising model (CDIM), capturing the simultaneous emergence of spin and polar order. Chalcogen vacancies act as sources of both W-centered magnetic moments and local electric dipoles, while Te substitution tunes the efficiencies $\alpha(x)$ and $\beta(x)$ with which each vacancy contributes to magnetic and polar activity. The fractions of active sites are
\begin{equation}
p_\mathrm{mag} = C_v^M \, \delta \, \alpha(x), \qquad
p_\mathrm{elec} = C_v^E \, \delta \, \beta(x),
\end{equation}
where $C_v^M$ and $C_v^E$ denote the number of magnetic moments and dipoles per vacancy. Differences between $C_v^M$ and $C_v^E$, and between $\alpha(x)$ and $\beta(x)$, originate from distinct microscopic mechanisms: magnetic moments arise from unpaired W electrons, whereas dipoles are generated by local lattice distortions. Additional factors such as spatial extent, lattice symmetry, bond stiffness, interaction range, and Te-induced orbital hybridization further differentiate the per-vacancy contributions, producing $C_v^M \neq C_v^E$ and $\alpha(x) \neq \beta(x)$. 

Percolation thresholds for magnetism and polarization therefore occur at different vacancy densities, giving rise to ferromagnetic–paraelectric, paramagnetic–ferroelectric, and multiferroic-like regions. The critical vacancy fractions required for long-range order are
\begin{equation}
\delta_c^\mathrm{mag}(x) = \frac{p_c}{C_v^M \, \alpha(x)}, \qquad
\delta_c^\mathrm{elec}(x) = \frac{p_c}{C_v^E \, \beta(x)},
\end{equation}
where $p_c$ is the lattice percolation threshold. The empirical separatrix approximates these thresholds, with $\delta_0 \approx \delta_c(x_0)$ at the optimal Te fraction $x_0$. Deviations from $x_0$ require additional vacancies to reach percolation, reflecting the nonlinear coupling of vacancy concentration and composition to the emergence of ferroic order.




\section{Summary and Conclusions}

Our results establish that ferroic behavior in W(Se,Te)$_2$ is primarily governed by chalcogen-site occupancy $\delta$, whereas Te fraction $x$ controls structural symmetry. This separation of roles is consistent with recent reports of room-temperature multiferroicity in Te-doped WSe$_2$~\cite{cardenas2023room}, where multiferroicity emerges for $20\%\!\lesssim\!\delta\!\lesssim\!38\%$. In contrast, we observe that piezoelectricity persists down to near-stoichiometric compositions ($\delta\!\approx\!2$–$4\%$), and that magnetism scales systematically with deviations from ideal chalcogen stoichiometry. 

We introduce a configurational phase diagram in $(x,\delta)$ space that unifies structural and electronic transitions. Expressing compositions equivalently through $(n_\mathrm{Te}/n_\mathrm{W},\, n_\mathrm{Se}/n_\mathrm{W})$ ratios provides a transparent representation of how defect chemistry and substitutional disorder determine ferroic responses. This framework is directly extensible to other transition-metal dichalcogenide alloys.

Structurally, $x$ acts as the effective order parameter for the $2H \rightarrow 1T_d$ transition between WSe$_2$ and WTe$_2$, primarily through lattice expansion and symmetry lowering (notably the increase of the $c$ parameter). In contrast, ferroic order emerges predominantly as a function of $\delta$. Low-$\delta$ compositions exhibit weak paramagnetism and piezoelectricity, whereas large $\delta$ stabilizes ferromagnetic and ferroelectric behavior, consistent with a defect-driven electronic instability mechanism.

The resulting hierarchy is therefore:
(i) $x$ controls crystallographic phase stability,
(ii) $\delta$ controls electronic instabilities leading to ferroicity,
(iii) multiferroic-like coexistence occurs when both magnetic and polar thresholds are exceeded within a single structural phase.

Although ferromagnetism and ferroelectricity have been separately reported in WSe$_2$ and WTe$_2$, it remains to be established whether nonstoichiometric WSe$_{2(1-\delta)}$ and WTe$_{2(1-\delta)}$ exhibit intrinsic defect-coupled multiferroicity. Finally, given the exfoliable nature of these layered systems, the dimensional evolution of vacancy-induced ferroicity toward the two-dimensional limit remains an open and experimentally accessible problem.
\section*{Acknowledgments}
The authors acknowledge the support of the X-ray diffraction and fluorescence laboratory of the Faculty of Science of Universidad de Los Andes. The authors acknowledge QuAnLab, the service laboratory network of the Department of Chemistry at the Universidad de los Andes, for access to Raman spectroscopy instrumentation and technical support. The authors acknowledge the instruments and scientific and technical assistance of the MicroCore Microscopy Core at the Universidad de Los Andes, a facility that is supported by the vicepresidency for research and creation. P.G-G. and E. R-R. gratefully acknowledge funding from the project ‘Ampliación del uso de la mecánica cuántica desde el punto de vista experimental y su relación con la teoría, generando desarrollos en tecnologías cuánticas útiles para metrología y computación cuántica a nivel nacional’, BPIN 2022000100133, from SGR of MINCIENCIAS, Gobierno de Colombia.
\appendix
\section{Compositional phase diagrams} \label{App_compositional_phase_diagram}

The general composition of the alloy family can be written as
\begin{equation}
\text{W}_{n_\mathrm{W}}\text{Se}_{n_\mathrm{Se}}\text{Te}_{n_\mathrm{Te}}
= \text{WSe}_{\frac{n_\mathrm{Se}}{n_\mathrm{W}}}
\text{Te}_{\frac{n_\mathrm{Te}}{n_\mathrm{W}}},
\end{equation}
where the two independent compositional variables are 
$\frac{n_\mathrm{Se}}{n_\mathrm{W}}$ and 
$\frac{n_\mathrm{Te}}{n_\mathrm{W}}$.

Defining the total chalcogen content as 
$n = n_\mathrm{Se} + n_\mathrm{Te}$, the measurable chalcogen-to-tungsten ratio becomes
\begin{equation}
\frac{n}{n_\mathrm{W}} 
= \frac{n_\mathrm{Se}}{n_\mathrm{W}} 
+ \frac{n_\mathrm{Te}}{n_\mathrm{W}}.
\end{equation}

Stoichiometric compositions satisfy
\begin{equation}
\frac{n}{n_\mathrm{W}} = 2,
\end{equation}
corresponding to the MX$_2$ condition of two chalcogen atoms per metal atom.

In the $(n_\mathrm{Te}/n_\mathrm{W},\, n_\mathrm{Se}/n_\mathrm{W})$ diagram (Fig.~\ref{Phase_Space}), these compositions lie along the straight line
\begin{equation}
\frac{n_\mathrm{Se}}{n_\mathrm{W}}
= -\frac{n_\mathrm{Te}}{n_\mathrm{W}} + 2.
\end{equation}

To account for deviations from ideal stoichiometry, the generalized expression
\begin{equation}
\text{W}[\text{Te}_x\text{Se}_{1-x}]_{2(1-\delta)}
\end{equation}
is adopted, with
\begin{equation}
\frac{n}{n_\mathrm{W}} = 2(1-\delta).
\end{equation}

Lines of constant $\delta$ (isostoichiometric lines) satisfy
\begin{equation}
\frac{n_\mathrm{Se}}{n_\mathrm{W}}
= -\frac{n_\mathrm{Te}}{n_\mathrm{W}} + 2(1-\delta),
\end{equation}
and appear as parallel straight lines with negative slope in the configurational diagram.

Lines of constant substitution fraction $x$ (isocompositional lines) satisfy
\begin{equation}
\frac{n_\mathrm{Se}}{n_\mathrm{W}}
= \left(\frac{1-x}{x}\right)
\frac{n_\mathrm{Te}}{n_\mathrm{W}},
\end{equation}
corresponding to radial trajectories emanating from the origin.

The $(x,\delta)$ parameterization therefore provides a compact and geometrically transparent description of the accessible compositional space, partitioning it into chalcogen-deficient ($\delta>0$), stoichiometric ($\delta=0$), and chalcogen-rich ($\delta<0$) regimes.
\section{Defect classification and formation-energy considerations}
\label{App_Defec_Hierarchy}
In solid solutions containing tellurium, achieving perfect stoichiometry is often challenging due to its tendency to segregate during sintering and crystal growth, a behavior arising from its distinct chemical characteristics, including comparatively weak reactivity with tungsten in W–Te systems \cite{dimitrov2020chemical}. Although chalcogen imbalances can be partially mitigated by controlling the synthesis atmosphere (e.g., using inert-gas environments), tellurium chemistry generally favors the formation of defects such as chalcogen vacancies, resulting in non-stoichiometric configurations. More generally, real crystals grown at finite temperatures contain intrinsic point defects with finite equilibrium concentrations determined by their formation energies \cite{libowitz1965nonstoichiometry}, and when a particular defect type is energetically favored, the resulting imbalance produces measurable deviations from ideal stoichiometry.

\subsection{Defect classification}

To clarify the microscopic origins of stoichiometric deviation, we consider the crystallographic point defects that can occur in transition metal dichalcogenides of the form MX$_2$ (M = transition metal, X = chalcogen).

Six fundamental types of isolated point defects are typically distinguished:
\textit{(1)} chalcogen vacancies [$\square_X$],
\textit{(2)} metal vacancies [$\square_M$],
\textit{(3)} chalcogen interstitials [$X_I$],
\textit{(4)} metal interstitials [$M_I$],
\textit{(5)} metal atoms on chalcogen sites [$M(X)$],
\textit{(6)} chalcogen atoms on metal sites [$X(M)$]
\cite{libowitz1962nonstoichiometry,libowitz1965nonstoichiometry,libowitz1967characterization,libowitz1969thermodynamic}.

In addition to isolated defects, higher-order configurations involving adjacent defect pairs—such as double metal ($\square_{MM}$), double chalcogen ($\square_{XX}$), and mixed ($\square_{MX}$) vacancies—may form. However, these complex defects generally exhibit higher formation energies and are therefore expected to have lower equilibrium concentrations \cite{haldar2015systematic}. Consequently, stoichiometric deviations are primarily governed by the relative populations of the six fundamental point defects.

Within this framework, compositions with $\delta > 0$ (i.e., $n_{\mathrm{Te}} + n_{\mathrm{Se}} < 2 n_{\mathrm{W}}$) correspond to chalcogen-deficient conditions and may involve $\square_X$, $M_I$, or $M(X)$ defects and their combinations. Conversely, compositions with $\delta < 0$ (i.e., $n_{\mathrm{Te}} + n_{\mathrm{Se}} > 2 n_{\mathrm{W}}$) correspond to chalcogen-rich conditions and may involve $\square_M$, $X_I$, or $X(M)$ defects.

\subsection{Defect energetics}

Although defects may be randomly distributed throughout the crystal, their equilibrium concentrations are governed by their formation energies. Within a thermodynamic framework, the probability of occurrence of a defect $\text{D}_i$ with formation energy $E_f^{(i)}$ follows a Boltzmann factor,
\begin{equation}
P(\text{D}_i) \propto \exp\left(-\frac{E_f^{(i)}}{k_B T_G}\right),
\end{equation}
where $k_B$ is the Boltzmann constant and $T_G$ is the crystal growth temperature \cite{libowitz1969thermodynamic,haldar2015systematic}. For bulk crystals synthesized by CVT, defect populations are assumed to reflect near-equilibrium concentrations established during growth.

\begin{figure}[htb!]
\centering
\includegraphics[width=0.485\textwidth]{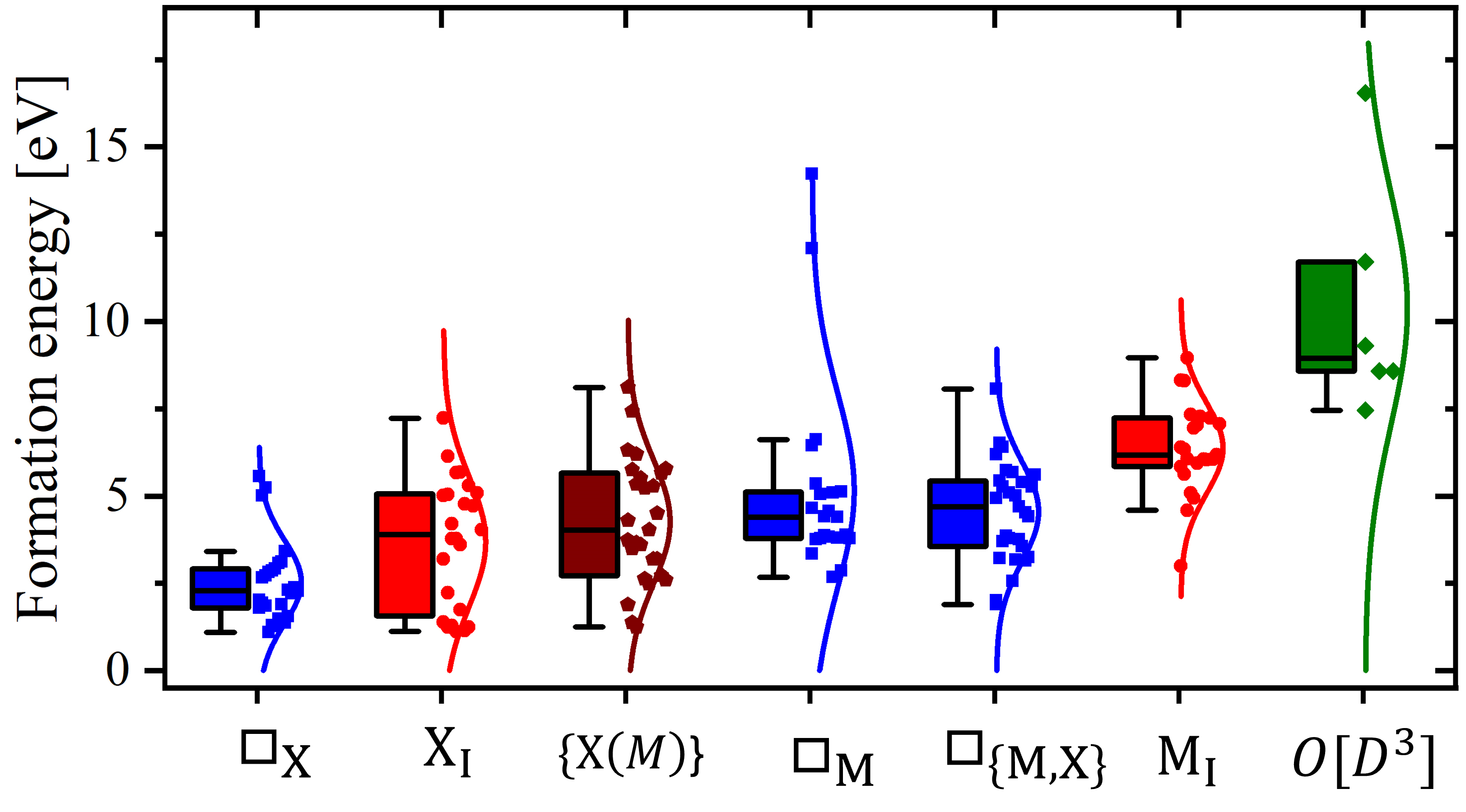}
\caption{
Comparison of the formation energies of various point defects in WSe$_2$ and WTe$_2$. Data were compiled from several theoretical studies based on \textit{ab initio} and density functional theory (DFT) calculations \cite{Haldar2015,akkoush2024,lin2015three,freire2022vacancy,kim2022experimental,zheng2019first,huang2022enhanced,yang2019electronic,han2019chemical}. The symbols $\{X(M)\}$, $\square_{\{X, M\}}$, and $O[D^3]$ represent, respectively: chalcogen/metal site substitutions ($X(M)$ and $M(X)$); double vacancies such as $\square_{MM}$, $\square_{XX}$, and $\square_{\{XM\}}$; and higher-order point defects, including configurations like $M(XX)$, $XX(M)$, $\square_{XXX}$, and $\square_{3XM}$.
}
\label{Formation_Energy}
\end{figure}

Numerous first-principles studies based on density functional theory (DFT) have reported formation energies for intrinsic defects in WSe$_2$ and WTe$_2$ under both chalcogen-rich and metal-rich conditions
\cite{Haldar2015,akkoush2024,lin2015three,freire2022vacancy,kim2022experimental,zheng2019first,huang2022enhanced,yang2019electronic,han2019chemical}. 
Figure~\ref{Formation_Energy} compiles representative formation energies reported for bulk and monolayer phases.

Across these studies, chalcogen vacancies ($\square_X$) and chalcogen interstitials ($X_I$) generally emerge among the lowest-formation-energy intrinsic defects under off-stoichiometric conditions. In contrast, double vacancies and higher-order complexes exhibit substantially higher formation energies and are therefore less probable in equilibrium. This energetic hierarchy suggests that deviations from ideal stoichiometry in WSe$_2$–WTe$_2$ alloys are most plausibly dominated by chalcogen vacancies in chalcogen-deficient regimes ($\delta>0$) and by interstitial-related defects in chalcogen-rich regimes ($\delta<0$).
\section{Atomic Radii}\label{Atomic_Radii}
Accurate modeling of transition metal dichalcogenides (TMDs) requires knowledge of different types of atomic radii shown in Table \ref{Radii_Table}. The \textit{covalent radius}$^a$ is defined as half the distance between two atoms connected by a covalent bond \cite{Cordero2008}. The \textit{van der Waals radius}$^b$ represents half the distance between two nonbonded atoms at their closest approach \cite{Bondi1964,batsanov2001van,alvarez2013cartography}. The \textit{metallic radius}$^c$ corresponds to half the distance between atoms in a metallic lattice \cite{Ashcroft1976}. Finally, the \textit{ionic radius}$^d$ describes the effective size of an atom in a given oxidation state and coordination environment \cite{Shannon1976}.

\begin{table}[h!]\label{Radii_Table}
\centering
\caption{Atomic radii of W, Se, and Te relevant for TMD modeling (Å): covalent (Cov), van der Waals (vdW), metallic (Met), and ionic (Ion). Coordination number CN=6 for the reported ionic radii.}
\small
\begin{tabular}{ccccc}
\hline
Element & Cov$^a$ & vdW$^b$ & Met$^c$ & Ion$^d$ \\
\hline
W  & 1.35 & 2.1 & 1.39 & 0.60 (W$^{6+}$) \\
Se & 1.20 & 1.90 & 1.21 & 1.98 (Se$^{2-}$) \\
Te & 1.38 & 2.06 & 1.40 & 2.21(Te$^{2-}$) \\
\hline
\end{tabular}
\end{table}

\section{Hexagonal Closed Packing}\label{App_hcp}
Under the close-packing approximation for atoms with an average radius $\overline{r}$, the relation $a = 2\overline{r}$ holds under the assumption of solid spheres in contact. For the 2H crystallographic phase, composed of two hexagonal layers in trigonal prismatic coordination, the $c$ lattice parameter is given by:
\begin{equation}\label{c_radius}
c = 10 \overline{r} \sqrt{\frac{2}{3}}.
\end{equation}
In alloys of the form $\text{W}_{n_\text{W}} \text{Se}_{n_\text{Se}} \text{Te}_{n_\text{Te}}$, the average atomic radius, in terms of the atomic radii of tungsten ($r_\text{W}$), selenium ($r_\text{Se}$), and tellurium ($r_\text{Te}$), is given by:
\begin{equation}
\label{Mean_atomic_radious}
\overline{r} = \frac{n_\text{Se} r_\text{Se} + n_\text{Te} r_\text{Te} + n_\text{W} r_\text{W}}{n_\text{Se} + n_\text{Te} + n_\text{W}} = \frac{\frac{n_\text{Se}}{n_\text{W}} r_\text{Se} + \frac{n_\text{Te}}{n_\text{W}} r_\text{Te} + r_\text{W}}{f(n) + 1},
\end{equation}
where $f(n) = \frac{n}{n_\text{W}}$, and $n = n_\text{Se} + n_\text{Te}$, as defined in Section~\ref{WSeTe_phase} in the body of the article.

By substituting the molar ratios from and the average atomic radius from Eq.~\ref{Mean_atomic_radious} into Eq.~\ref{c_radius}, the expression for the mean $c$ lattice parameter becomes:
\begin{equation}\label{c_equation}
c(x,\delta)=m(\delta)x+b(\delta).
\end{equation}
where
\begin{equation}
\begin{split}
m(\delta)&=10\sqrt{\frac{2}{3}}\left[ \frac{2(1-\delta)}{2(1-\delta)+1}(r_\text{Te}-r_\text{Se}) \right]\\ b(\delta)&=10\sqrt{\frac{2}{3}}\left[ \frac{2(1-\delta)}{2(1-\delta)+1} \left( r_\text{Se}+\frac{r_\text{W}}{2(1-\delta)}\right)\right].  
\end{split}
\end{equation}
For $\delta=0$ and $x=0$, it should be obtained the $c_0$ lattice parameter for pure 2H-WSe$_2$.
\setcounter{figure}{0} 
\makeatletter 
\renewcommand{\thefigure}{A\@arabic\c@figure}
\makeatother
\bibliography{Bibliography/biblio_Solid_solution_with_perfect_chalcogen_stoichiometry}

\end{document}